\documentclass[twocolumn,aps,showpacs,showkeys,bbm]{revtex4-1}
\usepackage[latin9]{inputenc}
\usepackage[active]{srcltx}
\usepackage{color}
\usepackage{array}
\usepackage{units}
\usepackage{textcomp}
\usepackage{epstopdf}
\usepackage{amsmath}
\usepackage{amssymb}
\usepackage{graphicx}
\usepackage{esint}
\usepackage[unicode=true,pdfusetitle,
 bookmarks=true,bookmarksnumbered=true,bookmarksopen=false,
 breaklinks=true,pdfborder={0 0 0},backref=false,colorlinks=true]
 {hyperref}

\makeatletter

\providecommand{\tabularnewline}{\\}

\usepackage{color}
\usepackage{bbm}

\makeatother

\begin{document}

\title{Pumping dynamics of nuclear spins in GaAs quantum wells}

\author{Raphael W. Mocek}

\email{raphael.mocek@tu-dortmund.de}

\affiliation{Experimental Physics III, TU Dortmund University, Otto-Hahn-Str.
4a, 44227 Dortmund,Germany}

\author{Danil O. Tolmachev}

\affiliation{Experimental Physics III, TU Dortmund University, Otto-Hahn-Str.
4a, 44227 Dortmund,Germany}

\author{Giovanni Cascio}

\affiliation{Experimental Physics III, TU Dortmund University, Otto-Hahn-Str.
4a, 44227 Dortmund,Germany}

\author{Dieter Suter}

\email{dieter.suter@tu-dortmund.de}

\affiliation{Experimental Physics III, TU Dortmund University, Otto-Hahn-Str.
4a, 44227 Dortmund,Germany}

\date{\today}
\begin{abstract}
Irradiating a semiconductor with circularly polarized light creates
spin-polarized charge carriers. If the material contains atoms with
non-zero nuclear spin, they interact with the electron spins via the
hyperfine coupling. Here, we consider GaAs/AlGaAs quantum wells, where
the conduction-band electron spins interact with three different types
of nuclear spins. The hyperfine interaction drives a transfer of spin
polarization to the nuclear spins, which therefore acquire a polarization
that is comparable to that of the electron spins. In this paper, we
analyze the dynamics of the optical pumping process in the presence
of an external magnetic field while irradiating a single quantum well
with a circularly polarized laser. We measure the time dependence
of the photoluminescence polarization to monitor the buildup of the
nuclear spin polarization and thus the average hyperfine interaction
acting on the electron spins. We present a simple model that adequately
describes the dynamics of this process and is in good agreement with
the experimental data. 
\end{abstract}

\maketitle

\section{Introduction}
\label{sec:Introduction} 
\begin{figure}[htbp]
\noindent \centering{}\centering \includegraphics[width=0.9\columnwidth]{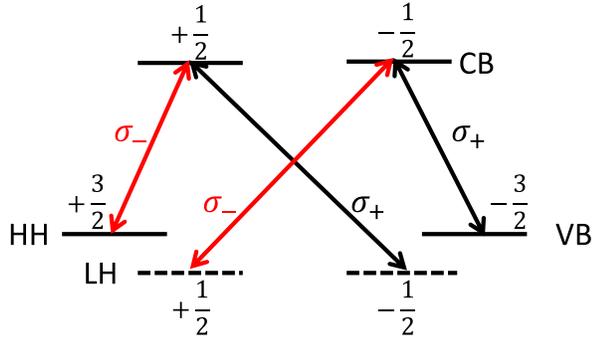}
\caption{Selection rules for the optical transitions between the valence-band
(VB) and the conduction-band (CB) of a semiconductor quantum well.
The confinement lifts the degeneracy of the hole states. (modified
after\,\cite{DYAKONOV1984})}
\label{fig:SelectionRules} 
\end{figure}

Optical pumping creates electrons and holes in semiconductor samples with spin polarizations far from equilibrium,
as shown by the level scheme and transition diagram of fig.\,\ref{fig:SelectionRules}.
Depending on the conduction- and valence-band states involved in the
optical transition, the polarization of the electron spins can reach
almost $100\%$\,\cite{Pfalz2005}.

The spin-polarization of the conduction-band electrons can be monitored
through the photoluminescence (PL) polarization\,\cite{ekimov1970optical,Parsons1969,Dzhioev}
\begin{equation}
DOP=\frac{I\left(\sigma_{+}\right)-I\left(\sigma_{-}\right)}{I\left(\sigma_{+}\right)+I\left(\sigma_{-}\right)},\label{eq:0.1}
\end{equation}
where $I\left(\sigma_{\pm}\right)$ is the intensity of right- or
left-circularly polarized PL. 

In many materials, and in particular in the GaAs quantum wells that
we consider in this work, the electrons are coupled to different nuclear
spins by the hyperfine interaction. Accordingly, the electron spin
orientation can be transferred to the nuclear spins\,\cite{Barrett1994,Pietra1996,Ekimov1972,Leifson1961}
in a process known as dynamic nuclear polarization (DNP)\,\cite{Lampel1968,DYAKONOV1984,Abragam1961}.
Conversely, the ensemble of polarized nuclear spins affect the evolution
of the electron spins. The overall effect can be summarized by an
effective nuclear magnetic field\,\cite{Paget1977,Lampel1968,Salis2001,Snelling1991,Kalevich1990,berkovits1973optical,d1974optical}.
This is used, e.g., in optically detected nuclear magnetic resonance\,\cite{Barrett1994,Salis2001,Cavenett1981,1382,5839,6064,6680}. 

The goal of this study is a detailed understanding of the buildup
of the nuclear spin polarization during optical pumping. For this
purpose, we rely mostly on measurements of the time dependence of
the PL polarization, from which we determine the buildup of the nuclear
magnetic field. The paper is organized as follows. In Sec.\,\ref{sec:Theory},
we describe a simple model of the spin dynamics during optical pumping.
In Sec.\,\ref{sec:Experimental Methods} we present the experimental
setup and the sample under investigation. Section\,\ref{sec:Optical pumping time dependence}
contains the experimental results for the time dependence of the optical
pumping process and compares them to the theoretical prediction. Section\,\ref{sub:dep_kappa}
summarizes the effect of the control parameters laser intensity and
optical detuning on the optical pumping dynamics. The paper ends with
a short discussion and conclusions.

\section{Theory}
\label{sec:Theory}

\subsection{Electron spin polarization}
\label{sec:Electron spin polarization} 

\begin{figure}[htbp]
\noindent \centering{}\centering \includegraphics[width=0.7\columnwidth]{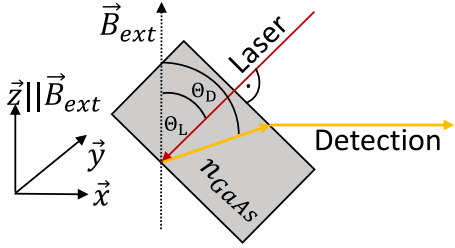}
\caption{Schematic overview of the chosen coordinate system and the relevant
angles. The laser light hits the sample perpendicularly.\textcolor{magenta}{{}
\label{fig:angles_gaas}}}
\end{figure}

The optical pumping process, as well as the optical detection couple
the photon angular momentum directly to the spin of the charge carriers.
We therefore start with the equation of motion for the spin density
operator $\rho$:

\begin{eqnarray}
\frac{\partial\rho}{\partial t} & = & -\frac{i}{\hbar}\left[\mathcal{H},\rho\right]-\Gamma_{R}\rho-\Gamma_{S}\left(\rho-\frac{Tr\left\{ \rho\right\} }{2}\mathbbm1\right)+\tilde{P}\label{eq:1}\\
\mathcal{\mathcal{H}} & = & \hbar\gamma_{e}\vec{B}\cdot\vec{S}\nonumber 
\end{eqnarray}
We use the spin operators defined as $S_{i}=\frac{1}{2}\sigma_{i}$
with $i\in\left[x,y,z\right]$, $\gamma_{e}$ is the gyromagnetic
ratio of the conduction electrons and $\mathbbm1$ is the two-dimensional
unity matrix. $\vec{B}$ is the magnetic field, $\Gamma_{R}$ describes
the recombination of the electrons from the conduction- to the valence-band
and $\Gamma_{S}$ the spin relaxation rate. The matrix $\tilde{P}$
describes the buildup of electron spin density by the absorption of
circularly polarized photons. In the coordinate system defined in
fig.\,\ref{fig:angles_gaas}, this matrix is 
\begin{equation}
\tilde{P}=P\left(\begin{array}{cc}
\cos^{2}\frac{\Theta_{L}}{2} & \frac{1}{2}\sin\Theta_{L}\\
\frac{1}{2}\sin\Theta_{L} & \sin^{2}\frac{\Theta_{L}}{2}
\end{array}\right).\label{eq:1.1}
\end{equation}
$P$ is the rate at which the optical pumping process generates electron
spin density in the conduction band. $\Theta_{L}$ is the angle between
the incident laser beam and the z-axis, as defined in fig.\,\ref{fig:angles_gaas}. 

If the nuclear spins are polarized, they collectively modify the effective
magnetic field acting on the electron spin. We therefore write the
total field $\vec{B}$ as the sum of the external field $\vec{B}_{ext}$
and the nuclear field $\vec{B}_{nuc}$. In our coordinate system,
both are oriented along the $z$-axis and we therefore write $B_{ext}+B_{nuc}$
for the $z-$component of the effective magnetic field. Under stationary
conditions, the expectation values of the three components of the
electron spin are 
\begin{eqnarray}
\left<S_{x}\right> & = & Tr\left\{ S_{x}\rho\right\} =\frac{P}{2\gamma_{e}}\frac{\Delta B\sin\Theta_{L}}{\left(B_{ext}+B_{nuc}\right)^{2}+\Delta B^{2}}\label{eq:7}\\
\left<S_{y}\right> & = & Tr\left\{ S_{y}\rho\right\} =\frac{P}{2\gamma_{e}}\frac{\left(B_{ext}+B_{nuc}\right)\sin\Theta_{L}}{\left(B_{ext}+B_{nuc}\right)^{2}+\Delta B^{2}}\nonumber \\
\left<S_{z}\right> & = & Tr\left\{ S_{z}\rho\right\} =\frac{P}{2\gamma_{e}}\frac{\cos\Theta_{L}}{\Delta B}.\nonumber 
\end{eqnarray}

\noindent Here the parameter $\Delta B$ is defined as $\Delta B=\frac{\hbar\left(\Gamma_{R}+\Gamma_{S}\right)}{\left|g^{*}\right|\mu_{B}}$
with $g^{*}$ as the g-factor of the conduction electrons\,\cite{Weisbuch1977,Hannak1995,PhysRevB.45.3922}
and $Tr\left\{ \rho\right\} =\frac{P}{\Gamma_{R}}$. In our case,
the direction of detection $\vec{e}$ is close to the x-axis 
\begin{equation}
\vec{e}=\left(\begin{array}{ccc}
\sin\Theta_{D}, & 0, & \cos\Theta_{D}\end{array}\right),\label{eq:4}
\end{equation}
where the angle $\theta_{D}$ is defined in fig.\,\ref{fig:angles_gaas}.

Experimentally, we measure the degree of photon polarization (see
eq.\,\eqref{eq:0.1}) in the direction $\vec{e}$ by dividing the
difference between the two intensities $I(\sigma_{\pm})$ by their
sum. The result is 
\begin{eqnarray}
S_{D} & = & \frac{1}{Tr\{\rho\}}\left(\sin\Theta_{D}\left<S_{x}\right>+\cos\Theta_{D}\left<S_{z}\right>\right)\nonumber \\
 & = & S_{0}\left(\cos\Theta_{L}\cos\Theta_{D}+\frac{\Delta B^{2}\sin\Theta_{L}\sin\Theta_{D}}{\Delta B^{2}+\left(B_{ext}+B_{nuc}\right)^{2}}\right).\label{eq:5.1}
\end{eqnarray}

\noindent Here $S_{0}=\frac{1}{2}\frac{\Gamma_{R}}{\Gamma_{R}+\Gamma_{S}}$
describes the equilibrium spin polarization in the absence of a magnetic
field. This signal, measured as a function of the external magnetic
field, is known as a (shifted) Hanle curve. It represents a Lorentzian,
with a maximum at $B_{ext}=-B_{nuc}$ and a width $\Delta B$.

\subsection{Nuclear spin polarization}
\label{sec:Dynamics of nuclear spin polarization} 
In this study,
we concentrate on the evolution of the nuclear spin polarization.
The equation of motion for the nuclear spin populations can be written
as 
\begin{flalign}
\frac{d}{dt}\left(\begin{array}{c}
p_{\uparrow}\\
p_{\downarrow}
\end{array}\right)=\left(\begin{array}{rr}
-\kappa s_{\downarrow}-\frac{1}{2T_{1}} & \kappa s_{\uparrow}+\frac{1}{2T_{1}}\\
\kappa s_{\downarrow}+\frac{1}{2T_{1}} & -\kappa s_{\uparrow}-\frac{1}{2T_{1}}
\end{array}\right)\left(\begin{array}{c}
p_{\uparrow}\\
p_{\downarrow}
\end{array}\right),\label{eq:10}
\end{flalign}
where $\kappa$ is the transfer rate at which electronic and nuclear
spins undergo mutual flip-flop transitions and $s_{\uparrow\downarrow}$
are the densities of the electron spins generated by the optical pumping
process. According to eq.\,\eqref{eq:7} they are 
\[
s_{\uparrow\downarrow}=\frac{P}{2\Gamma_{R}}\pm\left<S_{z}\right>.
\]

For quantum wells with dimensions of $\approx20$ nm, the recombination
rate of the electrons is $\Gamma_{R}\approx\unit[10^{9}]{s^{-1}}$\,\cite{Shi-Rong1994}.
For time-independent parameters, eq.\,\eqref{eq:10} can be solved
analytically. If the system is initially in thermal equilibrium, $p_{\uparrow\downarrow}\left(0\right)=0.5$,
the solution is 

\begin{equation}
p_{\uparrow\downarrow}\left(t\right)=\frac{1\pm\Delta p(t)}{2},
\end{equation}
where the population difference is 
\begin{equation}
\Delta p(t)=\Delta p_{\infty}\left(1-e^{-\left(\frac{1}{T_{1}}+\kappa\frac{P}{\Gamma_{R}}\right)t}\right)
\end{equation}
and it's equilibrium value
\begin{equation}
\Delta p_{\infty}=\left<S_{z}\right>\frac{2\kappa T_{1}\Gamma_{R}}{\Gamma_{R}+\kappa T_{1}P}.\label{eq:deltap}
\end{equation}

\noindent The nuclear spin polarization can be measured through its
effect on the electron spin, via the effective nuclear field 
\begin{eqnarray}
B_{nuc}\left(t\right) & = & B_{max}\Delta p\left(t\right)\nonumber \\
 & = & B_{max}\Delta p_{\infty}\left(1-e^{-\left(\frac{1}{T_{1}}+\kappa\frac{P}{\Gamma_{R}}\right)t}\right).\label{eq:bnuc}
\end{eqnarray}
According to eq.\,\eqref{eq:5.1}, $B_{nuc}$ is given by the maximum
of the Hanle curve. To measure the time dependence of the populations
of the nuclear spin, we therefore measure the Hanle curves for different
pumping times.

\section{Experimental}
\label{sec:Experimental Methods} 

\noindent \label{sec:Sample-1} The sample used for this investigation
was grown by molecular beam epitaxy on a Te-doped GaAs substrate.
It consists of $13$ undoped $\text{GaAs/Al}_{0.35}\text{Ga}_{0.65}\text{As}$
quantum wells with thicknesses $d$ ranging from 2.8 to $\unit[39.3]{nm}$\,\cite{Eshlaghi08}.

\noindent 
\begin{figure}[htbp]
\noindent \centering{}\centering \includegraphics[width=1\columnwidth]{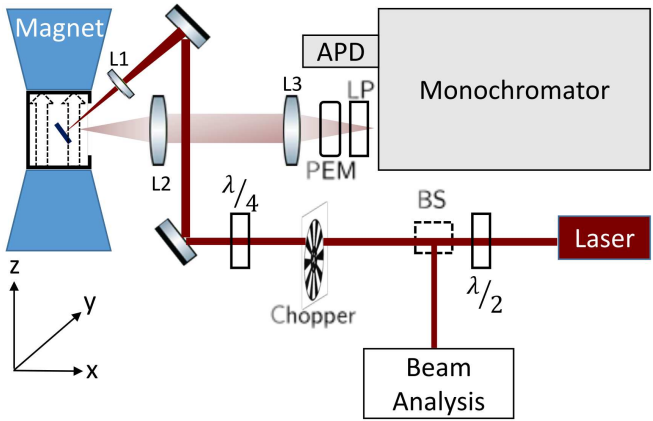}
\caption{Experimental setup of the optical pumping experiment. L1, L2 and L3:
lenses, LP: linear polarizer, PEM: photo elastic modulator, APD: avalanche
photo diode, BS: beam splitter, $\lambda/2$, $\lambda/4$ : retardation
plates.}
\label{fig:Aufbau} 
\end{figure}

\noindent Figure\,\ref{fig:Aufbau} shows a schematic representation
of the experimental setup. For the optical excitation we use a semiconductor
laser (Toptica DLC DL PRO), which covers the wavelength range of $\lambda_{exc}=\unit[799-812]{nm}$.
The magnetic field is created by a resistive electromagnet (Bruker)
with a range of $B_{ext}=\unit[0]{T}-\unit[1.4]{T}$. The sample is
mounted on the cold finger of a home-built flow-cryostat and kept
at temperatures of $T\approx\unit[4.7\pm0.3]{K}$. The laser beam
analysis includes a spectrometer (APE waveScan USB) for monitoring
the laser wavelength and a photodiode to monitor the laser power.
The PL is passed through a monochromator (Spex $1704$) and a photo-elastic
modulator (Hinds Instruments PEM 90) and detected with an avalanche
photodiode (APD, Hamamatsu $5640$). Two lock-in amplifiers (Stanford
Research SR830 DSP) are then used to measure the total PL power $I_{\Sigma}=I\left(\sigma_{+}\right)+I\left(\sigma_{-}\right)$
and the difference between right- and left-circularly polarized light
$I_{\Delta}=I\left(\sigma_{+}\right)-I\left(\sigma_{-}\right)$. 

\label{sec:Hanle curves} The optical pumping process took place in
constant external magnetic fields of $B_{ext}=\unit[0.3]{T}$ or $\unit[1]{T}$.
We monitored the buildup of the nuclear spin polarization by measuring
Hanle curves\,\cite{Hanle,DYAKONOV1984} as a function of the pumping
time. For the Hanle curves, the magnetic field was scanned from $B_{ext}$
either upward or downward, depending on the displacement of the Hanle
curve. The time for measuring a Hanle curve was about $\unit[10]{s}$,
short compared to the duration of the optical pumping. During the
Hanle measurements, the laser power was reduced to $P_{L}\approx\unit[2]{mW}$,
to minimize the optical pumping effects. The angles defined in fig.\,\ref{fig:angles_gaas}
were $\Theta_{D}=81^{\circ}$ and $\Theta_{L}=78^{\circ}$
for all experiments.

\section{Results}
\label{sec:Results}

\subsection{Nuclear field buildup}
\label{sec:Optical pumping time dependence} 

\begin{figure}[htbp]
\noindent \centering{}\includegraphics[width=0.9\columnwidth]{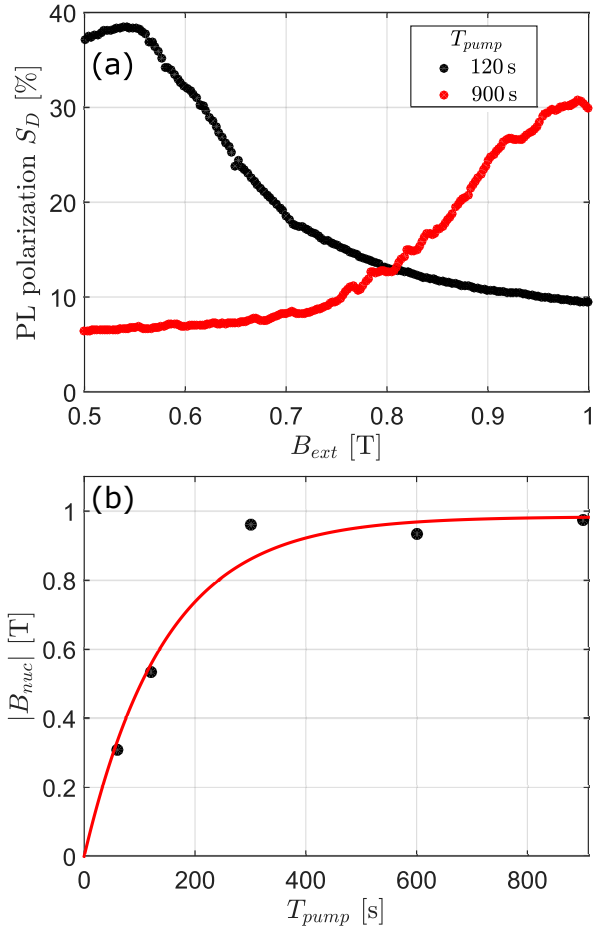}\caption{Time dependence of the effective nuclear field $\left|B_{nuc}\right|$:
(a) Hanle curves measured after optical pumping periods of $T_{pump}=\unit[120]{s}$
and $\unit[900]{s}$. (b) Evolution of the nuclear field during the
optical pumping. The solid red line is the fit result using eq.\,\eqref{eq:bnuc}
and the filled circles represent the experimental values. \label{fig:Kap4_Hanle_Tpump}}
\end{figure}

The main goal of these experiments was a quantitative understanding
of the process that generates the nuclear spin polarization. For this
purpose, we performed a set of measurements that consisted of a pumping
period $T_{pump}$ during which the sample was irradiated with circularly
polarized light in a constant magnetic field. Immediately after this
pumping period, we performed a rapid scan of the magnetic field to
measure a Hanle curve. According to eq.\,\eqref{eq:7}, the maxima
of these curves correspond to the effective nuclear field $\left|B_{nuc}\right|$
and can therefore be used as a probe of the nuclear spin polarization.
Measured Hanle curves after two different times $T_{pump}$ are shown
in fig.\,\ref{fig:Kap4_Hanle_Tpump}\,(a). The experimental parameters
for these measurements are $B_{ext}=\unit[1]{T}$ and laser power
$P_{L}=\unit[49]{mW}$. The monochromator was set to the maximum of
the PL line of the $d=\unit[19.7]{nm}$ quantum well and the optical
detuning of the laser beam was $\Delta\lambda=\lambda_{det}-\lambda_{exc}=\unit[811.6]{nm}-\unit[811.3]{nm}=\unit[0.3]{nm}$. 

Figure\,\ref{fig:Kap4_Hanle_Tpump}\,(b) shows the buildup of the
effective nuclear field $\left|B_{nuc}\left(t\right)\right|$. It
compares the experimental data with the theoretical curve calculated
from eq.\,\eqref{eq:bnuc}. For the parameters, we used a spin-lattice
relaxation time of $T_{1}=\unit[596\pm160]{s}$, which we measured
independently, and is comparable to literature values for similar
systems \,\cite{Berg1990}. From the fitted curve and our experimental
parameters, we calculated $P=\unit[7.7\cdot10^{13}]{\frac{\#e^{-}}{s\cdot\mu m^{3}}}$
. Table \ref{Tab:1-1} shows the other relevant parameters determined
from these curves.

\begin{table}[htbp]
\noindent \begin{centering}
\begin{tabular}{>{\centering}p{2.5cm}c>{\centering}p{2cm}}
$B_{max}\unit{\left[T\right]}$ & $\Delta p_{\infty}\unit{\left[\%\right]}$  & $\kappa\unit{\left[\frac{\mu m^{3}}{s\cdot\#e^{-}}\right]}$\tabularnewline
\hline 
$29$ & $3.4$ & $6.8\cdot10^{-8}$\tabularnewline
\hline 
\end{tabular}
\par\end{centering}
\caption{Fit parameters for the data shown in fig.\,\ref{fig:Kap4_Hanle_Tpump}\,(b).}
\label{Tab:1-1}
\end{table}

The fit-result for the maximum field $B_{max}$ is significantly larger
than some values from the literature \cite{DYAKONOV1984,Paget1977}. 

\begin{figure}[htbp]
\noindent \begin{centering}
\includegraphics[width=0.9\columnwidth]{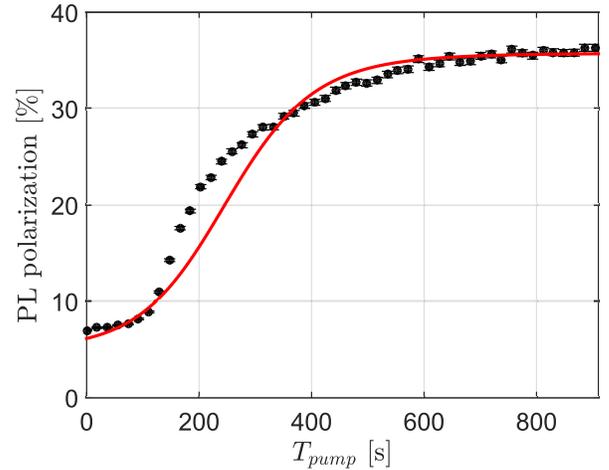}
\par\end{centering}
\centering 
\noindent \centering{}\caption{Time dependence of the optical pumping process: measured PL polarization
under optical pumping conditions with $B_{ext}=\unit[1]{T}$, $P_{L}=\unit[49]{mW}$
and $\Delta\lambda=\unit[0.3]{nm}$. The solid red line was calculated
with the parameters of table \ref{Tab:1-1} using eq.\,\eqref{eq:5.1}
and \eqref{eq:bnuc}.\label{fig:Results_TimeDep-1}}
\end{figure}

The buildup of the nuclear spin polarization can also be monitored
through the time dependence of the PL polarization during the optical
pumping process, as shown in fig.\,\ref{fig:Results_TimeDep-1}.
The theoretical curve was calculated from eqs.\,\eqref{eq:5.1} and
\eqref{eq:bnuc} with the parameters of tab.\,\ref{Tab:1-1}.

\subsection{Dependence on the laser intensity}
\label{sub:dep_kappa}

The parameter $P$ introduced in eq.\,\eqref{eq:1.1} describes the
rate at which electron spin density is created in the material. Over
some range, we therefore expect that $P$ should be proportional to
the laser power $P_{L}$. Since only absorbed photons generate electron
spins, the rate should also depend on the absorption probability of
the photons and reach a maximum at the optical resonance. The influence
of optical detuning is discussed in Sec.\,\ref{sec:Excitation wavelength}.

We examined the dependence of the pumping dynamics on the laser power
by performing a series of measurements with increasing laser intensity.
After pumping times $T_{pump}=\left[\unit[30]{s},\thinspace\unit[60]{s},\thinspace\unit[180]{s},\thinspace\unit[300]{s},\thinspace\unit[600]{s}\right]$,
we measured Hanle curves to monitor the evolution of the effective
nuclear magnetic field $\left|B_{nuc}\right|$. We repeated this procedure
with laser powers of $P_{L}=\left[\unit[20]{mW},\thinspace\unit[25]{mW},\thinspace\unit[30]{mW},\thinspace\unit[35]{mW},\thinspace\unit[40]{mW}\right]$.
Further experimental parameters were $B_{ext}=\unit[0.3]{T}$ and
$\Delta\lambda=\unit[0.5]{nm}$. The detection wavelength was $\lambda_{det}=\unit[811.6]{nm}$,
which corresponds to the maximum of the PL line of the $d=\unit[19.7]{nm}$
quantum well.

\begin{figure}[htbp]
\noindent \centering{}\centering \includegraphics[width=1\columnwidth]{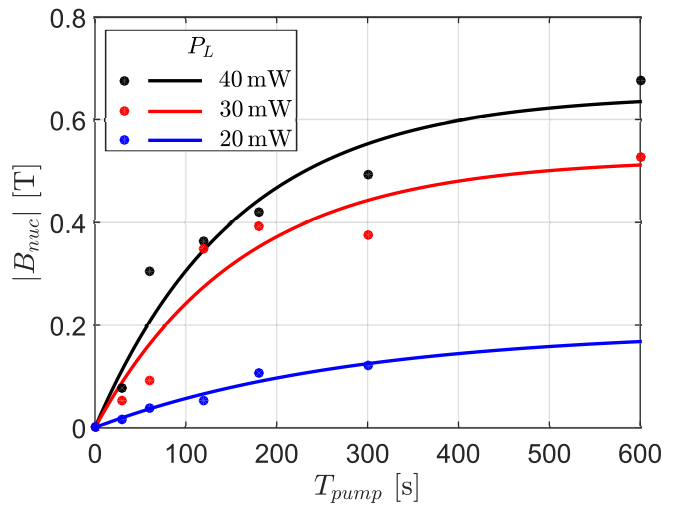}
\caption{Evolution of the effective nuclear magnetic field $\left|B_{nuc}\right|$
for different laser powers $P_{L}$. The circles mark the experimental
data while the solid lines are the result of the fit using eq.\,\eqref{eq:bnuc}
and $B_{max}=\unit[5.3]{T}$, $\kappa=\unit[6.3\cdot10^{-8}]{\frac{\mu m^{3}}{s\cdot\#e^{-}}}$
as fixed parameters. }
\label{fig:PL_pol_excitation} 
\end{figure}

Figure\,\ref{fig:PL_pol_excitation} shows the buildup of the effective
nuclear magnetic field $\left|B_{nuc}\left(t\right)\right|$ for different
laser intensities. The experimental data, which are shown as circles,
are the maxima of the Hanle curves taken after each optical pumping
period $T_{pump}$. We used $B_{max}=\unit[29]{T}$ and $\kappa=\unit[6.8\cdot10^{-8}]{\frac{\mu m^{3}}{s\cdot\#e^{-}}}$
as fixed parameters obtained by the measurement presented in Sec.\,\ref{sec:Optical pumping time dependence}
and eq.\,\eqref{eq:bnuc} to fit the experimental data shown in fig.\,\ref{fig:PL_pol_excitation}.
The resulting $\left|B_{nuc}\left(t\right)\right|$ are shown as solid
curves. 

\begin{figure}[htbp]
\noindent \begin{centering}
\includegraphics[width=1\columnwidth]{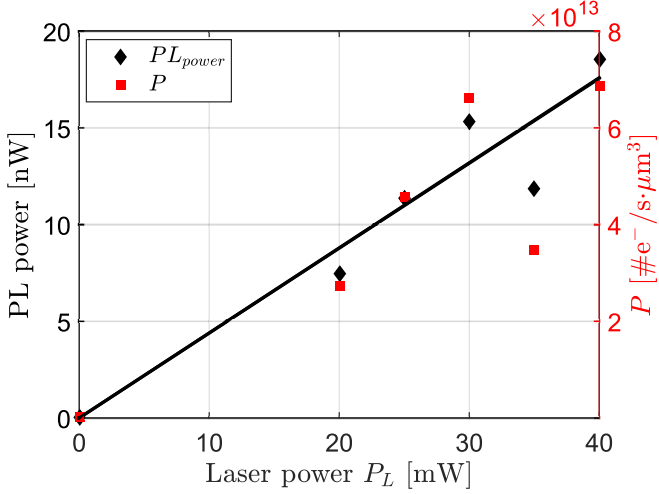}
\par\end{centering}
\caption{Rate of change in electron spin density $P$ together with the PL
light power as a function of the laser power $P_{L}$ . The solid
line is the result of the linear fit.\label{fig:kappa_Bnuc}}
\end{figure}

From the observed buildup-rate of $\left|B_{nuc}\right|$, we calculated
the rate $P$ at which electron-spin density is generated by the pumping
process. Figure\,\ref{fig:kappa_Bnuc} shows the resulting rates
$P$ as a function of the laser power $P_{L}$. The rate at which
electrons are generated in the conduction-band should also be reflected
in the PL power, which we measured independently. As shown in fig.\,\ref{fig:kappa_Bnuc},
both quantities are roughly proportional to the laser power, with
proportionality factors $m_{P}=\unit[\left(1.7\pm0.5\right)\cdot10^{12}]{\frac{\#e^{-}}{s\cdot\mu m^{3}\cdot mW}}$
and $m_{PL_{power}}=\unit[0.4\pm0.1]{\frac{nW}{mW}}$, respectively.

\subsection{Optical detuning}
\label{sec:Excitation wavelength} 

The rate $P$ at which electron-spins are generated depends also on
the frequency of the laser with respect to the resonance frequency
of the quantum well. We measured the time dependence of the PL polarization
under optical pumping conditions for different optical detunings $\Delta\lambda=\unit[[0.7\,...\,2.1]{]}$\,nm
relative to the peak wavelength $\lambda_{det}=\unit[811.6]{nm}$
of the $d=\unit[19.7]{nm}$ quantum well. The experimental parameters
were $B_{ext}=\unit[0.3]{T}$ and $P_{L}=\unit[47]{mW}$. 

\begin{figure}[htbp]
\noindent \centering{}\includegraphics[width=1\columnwidth]{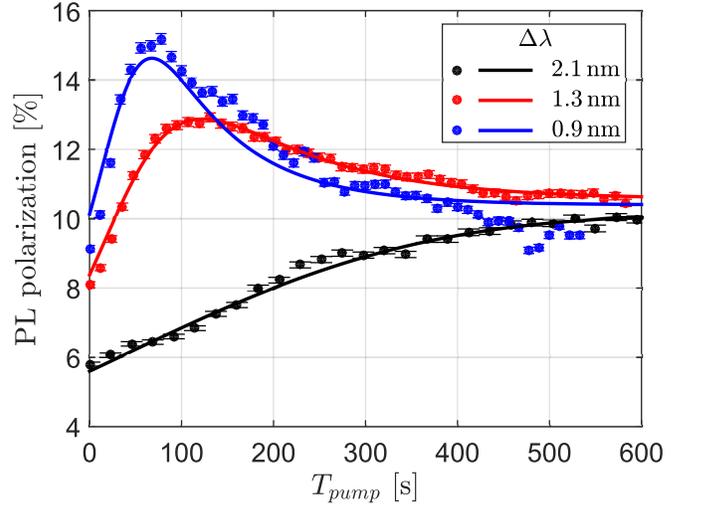}\caption{Time dependence of the optical pumping process: measured PL polarization
under optical pumping conditions with $B_{ext}=\unit[0.3]{T}$, $P_{L}=\unit[47]{mW}$
and $T_{pump}=\unit[600]{s}$ for different optical detunings $\Delta\lambda$.
The solid lines are the fit results based on eq.\,\eqref{eq:bnuc},\eqref{eq:5.1}.\label{fig:detuning_timesig}}
\end{figure}

Figure\,\ref{fig:detuning_timesig} shows some of the measured curves.
The experimental data are compared to the theoretical expectations
calculated from eq.\,\eqref{eq:5.1} and \eqref{eq:bnuc}, using
the experimental parameters of our system and the fit parameters given
in tab.\,\ref{Tab:1-1}. The electron-spin density generation rate
$P$ was adjusted for the curves to fit the experimental data.

\begin{figure}[htbp]
\noindent \begin{centering}
\includegraphics[width=1\columnwidth]{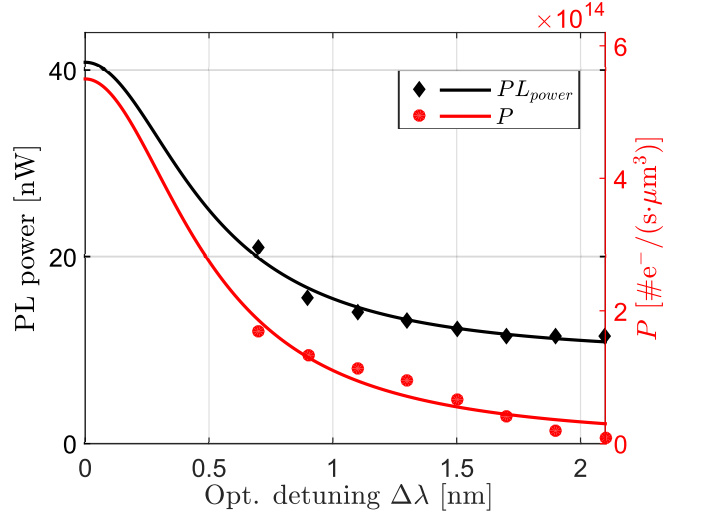}
\par\end{centering}
\caption{Detuning dependence of the rate of change in electron spin density
$P$: the circles mark the fit-parameters $P$ and the solid red line
is the result of the Lorentzian fit function eq.\,\eqref{eq:lorentz-1}.
Detuning dependence of the PL light power: the diamonds mark the measured
PL light power and the solid black line is given by eq.\,\eqref{eq:lorentz_PL-1}.
\label{fig:detuning_parameters}}
\end{figure}

\noindent Figure\,\ref{fig:detuning_parameters} shows the resulting
fit-parameters $P$ as red circles for the complete set of measurements
as a function of the laser detuning $\Delta\lambda.$ We compare the
experimental data points to a Lorentzian, 
\begin{equation}
P\left(\Delta\lambda\right)=\frac{a_{1}}{\left(\frac{\Delta\lambda}{HWHM}\right)^{2}+1}.\label{eq:lorentz-1}
\end{equation}

\noindent Fitting the data marked as circles in fig.\,\ref{fig:detuning_parameters}
to eq.\,\eqref{eq:lorentz-1} results in $a_{1}=\unit[\left(5.5\pm0.9\right)\cdot10^{15}]{\frac{\#e^{-}}{s\cdot\mu m^{3}\cdot nm}}$
and a half width at half maximum $HWHM=\unit[\left(0.5\pm0.1\right)]{nm}$. 

The number of photons absorbed by the QW should also be reflected
in the rate of emitted PL. We therefore also measured the average
PL power as a function of the detuning $\Delta\lambda$. The results
are shown in fig.\,\ref{fig:detuning_parameters} as black diamonds.
We compare them to the theoretically expected behavior of a Lorentzian
line similar to eq.\,\eqref{eq:lorentz-1}, but with an additional
offset $PL_{off}$ reflecting the effect of non-resonant excitation:
\begin{equation}
PL_{power}\left(\Delta\lambda\right)=\frac{a_{2}}{\left(\frac{\Delta\lambda}{0.5\,\mathrm{nm}}\right)^{2}+1}+PL_{off}.\label{eq:lorentz_PL-1}
\end{equation}
The fit result is shown in fig.\,\ref{fig:detuning_parameters} as
a solid black line using the fit-parameters $PL_{power}\left(\Delta\lambda=0\right)\unit[\approx40.8]{nW}$,
$a_{2}=\unit[31.7\pm2.1]{nW}$ and $PL_{off}=\unit[9.2\pm0.8]{nW}$.
The value of $PL_{off}$ was also measured independently by exciting
the system with a blue laser with $\lambda_{exc}=\unit[406]{nm}$
and a laser power of $P_{L}=\unit[10]{mW}$. We obtained a PL power
that was compatible with the value given above and found no significant
PL polarization, which is consistent with the above assumption that
the non-resonant processes should not contribute to the spin density\,\cite{6064}.

\subsection{Influence of the laser beam cross section}

\begin{figure}[htbp]
\noindent \begin{centering}
\centering \includegraphics[width=1\columnwidth]{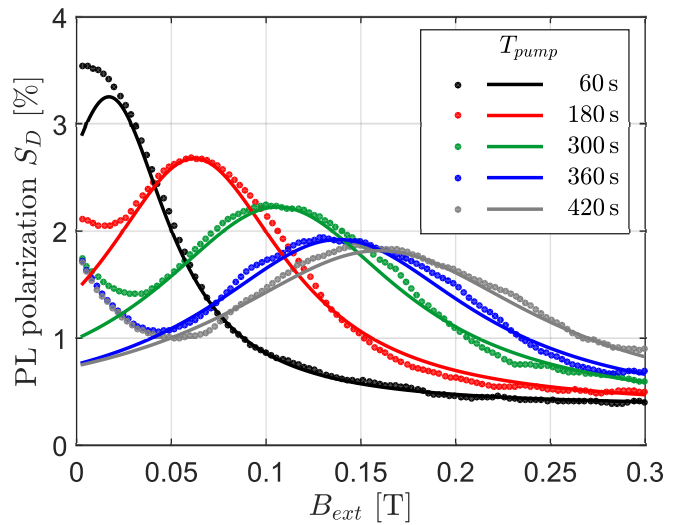}
\par\end{centering}
\caption{Shifted Hanle curves: measured data and calculated Hanle curves based
on the fit-parameters shown in fig.\,\ref{fig:Kap4_SC_Param}. \label{fig:hanle_meas_300mT_817nm}}
\end{figure}

Figure\,\ref{fig:hanle_meas_300mT_817nm} shows a representative
series of Hanle curves measured for different pumping times $T_{pump}$
in an external field $B_{ext}=\unit[0.3]{T}$ with a laser power of
$P_{L}=\unit[39]{mW}$. 
Similar to the data shown above, they show
a buildup of nuclear spin polarization, manifested as a shift of the
maximum towards higher fields. Compared to the data shown above, the
monochromator was set to the PL maximum of the $d=\unit[39.1]{nm}$
quantum well, and the optical detuning of the laser was $\Delta\lambda=\unit[7]{nm}$,
which resulted in correspondingly lower PL polarization. We used smaller
increments for $T_{pump}$ in order to achieve higher temporal resolution
of the pumping process. Using eq.\,\eqref{eq:5.1}, we fitted each curve separately to the
theoretical shape, adding an offset of $\unit[0.3]{\%}$, which may
be due to stray light. As shown in fig.\,\eqref{fig:hanle_meas_300mT_817nm},
we find the expected increase of the nuclear field with the pumping
time. In addition, the curves broaden and the maximum of the polarization
decreases with increasing pumping time. To understand these additional effects, we consider the intensity
distribution over the laser beam cross section, which we assume to
be Gaussian. As a result of this distribution, different sample regions
experience different intensities resulting in different pumping rates.
To describe this effect, we write the laser intensity as a function
of the distance $r$ from the center of the beam as 
\begin{equation}
I\left(r\right)=I_{0}\cdot e^{-\frac{r}{2\sigma_{0}^{2}}},\label{eq:13}
\end{equation}
with the maximum intensity 
\begin{equation}
I_{0}=\frac{P_{L}}{2\pi\sigma_{0}^{2}}.\label{eq:14}
\end{equation}
For a laser power $P_{L}=\unit[39]{mW}$ and beam width $\sigma_{0}=\unit[88]{\mu m}$,
we obtain $I_{0}=0.\unit[8]{\frac{\mu W}{\mu m^{2}}}$. For a numerical simulation of the observed Hanle curves, we calculated
the pumping dynamics and resulting Hanle curves $h\left(B_{ext},I\left(r\right)\right)$
for annular segments of the laser beam using eq.\,\eqref{eq:5.1}.
The individual curves were then weighted with the PL power $I\left(r\right)r\,dr$
emitted by each ring. The resulting calculated Hanle curve is 
\begin{equation}
S_{C}=\frac{\int dr\,h\left(B_{ext},I(r)\right)\,r\,I(r)}{\int dr\,r\,I(r)}.
\end{equation}
\begin{figure}[htbp]
\noindent \centering{}\includegraphics[width=1\columnwidth]{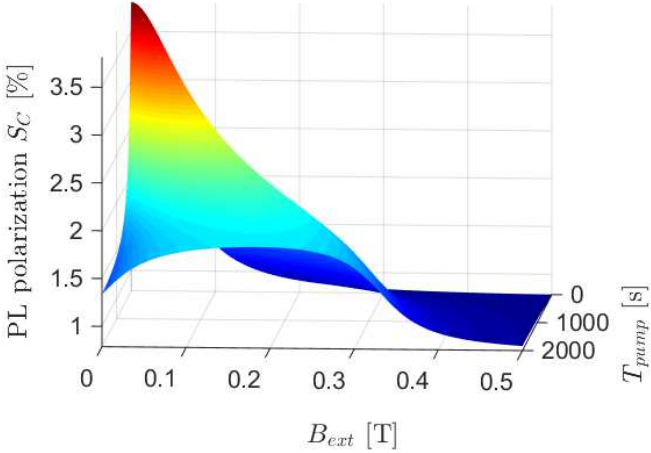}\caption{Calculated Hanle curves $S_{C}$ for increasing times $T_{pump}=\unit[[0...2000]{]\,}$s.
\label{fig:3D_SC}}
\end{figure}
Figure\,\ref{fig:3D_SC} shows the calculated Hanle curves $S_{C}$
for increasing times $T_{pump}=\unit[[0...2000]{]}{s}$. The saturation
of the nuclear field $B_{nuc}$ as well as the broadening of the curves
and the decrease in the maximal degree of PL polarization are clearly
visible.
\begin{figure}[htbp]
\noindent \centering{}\includegraphics[width=1\columnwidth]{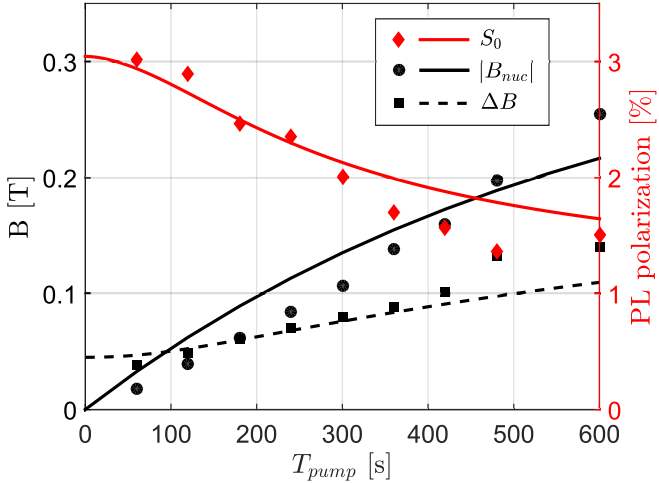}\caption{Comparison between the experimental parameters $\left|B_{nuc}\right|$,
$\Delta B$ and $S_{0}$ obtained by fitting the Hanle curves shown
in fig.\,\ref{fig:hanle_meas_300mT_817nm}\,(a) using eq.\,\eqref{eq:5.1}
with the result of the simulation.\label{fig:Kap4_SC_Param}}
\end{figure}
\noindent 
Figure\,\ref{fig:Kap4_SC_Param} compares the predictions
from this simple model with the experimental Hanle curves. The decreasing
maximal degree of PL polarization and the broadening of the curves,
represented by $\Delta B$, as well as the shift of the maxima, $\left|B_{nuc}\right|$,
are qualitatively well described by the theoretical model. 
\begin{figure}[htbp]
\noindent \begin{centering}
\includegraphics[width=1\columnwidth]{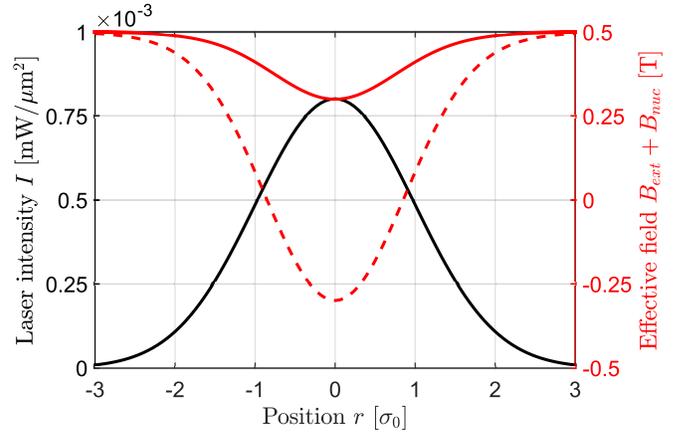}
\par\end{centering}
\caption{Laser intensity and effective field $B_{ext}+B_{nuc}$ with $B_{ext}=\unit[0.5]{T}$
as a function of the position in the laser beam in units of $\sigma_{0}=\unit[88]{\mu m}$.
The solid red line is calculated for a pumping time, $T_{pump}=\unit[36]{s}$
and the dashed red line for $T_{pump}=\unit[271]{s}$. \label{fig:effective_B_Position_r}}
\end{figure}
\noindent Figure\,\ref{fig:effective_B_Position_r} summarizes this
model in a different way: it shows the variation of the effective
field $B_{ext}+B_{nuc}$ over the laser beam cross section. The dependence
of the effective field on the position $r$ is calculated for two
different pumping times, $T_{pump}=\unit[36]{s}$ (solid red line)
and $T_{pump}=\unit[271]{s}$ (dashed red line) with an external field
of $B_{ext}=\unit[0.5]{T}$, and a laser power of $P_{L}=\unit[49]{mW}$.

\section{Discussion and Conclusion}
\label{sec:Discussion}

As we have shown, our experimental setup allows one to measure the
optical pumping dynamics of nuclear spins in GaAs quantum wells. We
presented a simple theoretical model that describes the buildup of
the nuclear spin polarization and therefore of the nuclear field in
a quantitative way and agrees with the experimental data, within the
experimental uncertainties. This model starts with the generation
of electron spins by the optical pumping process. The relevant rate
of electron spin density production is proportional to the laser power
$P_{L}$ and decreases with increasing optical detuning $\Delta\lambda$.
The spin polarization is then transferred from the electron spins
to the nuclear spins, and we also determined the rate constant for
this process. The experimental data show some significant deviations
from the simple model, which could be explained quantitatively by
taking into account that the laser beam does not illuminate the sample
homogeneously, but with a roughly Gaussian profile. These results
can be used to prepare the nuclear spin system e.g. for optically
detected nuclear magnetic resonance experiments. 

\begin{acknowledgments}
We gratefully acknowledge the support by the International Collaborative
Research Centre TRR 160 \textquotedblleft Coherent manipulation of
interacting spin excitations in tailored semiconductors,\textquotedblright{}
funded by the Deutsche Forschungsgemeinschaft. 
\end{acknowledgments}
\newpage

\bibliographystyle{apsrev}

\end{document}